\documentclass[prd,aps,showpacs,reprint,preprintnumbers,amsmath,amssymb,superscriptaddress,longbibliography,nofootinbib,floatfix]{revtex4-2}

\usepackage[utf8]{inputenc}
\usepackage{amsmath}
\usepackage{amssymb} 
\usepackage{enumerate}
\usepackage{multirow}
\usepackage{braket}
\usepackage[toc,page]{appendix}
\usepackage{url}
\usepackage{graphicx}
\usepackage[pdftex,bookmarks,linktocpage,pdfpagelabels,plainpages=false,hyperfigures,linkcolor=blue,citecolor=blue]{hyperref} 
\hypersetup{colorlinks=true}
\usepackage{adjustbox}
\usepackage{mathrsfs,amssymb}  
\usepackage{cancel}
\usepackage[normalem]{ulem}
\usepackage{cleveref}
\usepackage{tikz}
\usetikzlibrary{decorations.markings}
\usetikzlibrary{positioning}
\usetikzlibrary{shapes}
\usetikzlibrary{calc}

\newcommand{\beq}{\begin{equation}}
\newcommand{\eeq}{\end{equation}}

\newcommand{\bea}{\begin{eqnarray}}
\newcommand{\eea}{\end{eqnarray}}

\begin{document}

\title{Quantum similarity learning for anomaly detection} 
\author{A.\,Hammad}
\email{hamed@post.kek.jp}
\affiliation{Theory Center, IPNS, KEK, 1-1 Oho, Tsukuba, Ibaraki 305-0801, Japan.}

\author{Mihoko M. \,Nojiri}
\email{nojiri@post.kek.jp }
\affiliation{Theory Center, IPNS, KEK, 1-1 Oho, Tsukuba, Ibaraki 305-0801, Japan.}
\affiliation{The Graduate University of Advanced Studies (Sokendai), 1-1 Oho, Tsukuba, Ibaraki 305-0801, Japan.}
\affiliation{Kavli IPMU (WPI), University of Tokyo, 5-1-5 Kashiwanoha, Kashiwa, Chiba 277-8583, Japan.}

\author{Masahito \,Yamazaki}
\email{masahito.yamazaki@ipmu.jp}

\affiliation{Kavli IPMU (WPI), University of Tokyo, 5-1-5 Kashiwanoha, Kashiwa, Chiba 277-8583, Japan.}
\affiliation{Department of Physics, University of Tokyo, Hongo, Tokyo 113-0033, Japan}
\affiliation{Center for Data-Driven Discovery, Kavli IPMU,  University of Tokyo, Kashiwa, Chiba 277-8583, Japan}
\affiliation{Trans-Scale Quantum Science Institute, University of Tokyo, Hongo Tokyo 113-0033, Japan.}

\date{\today}

\begin{abstract}

Anomaly detection is a vital technique for exploring signatures of new physics Beyond the Standard Model (BSM) at the Large Hadron Collider (LHC). The vast number of collisions generated by the LHC demands sophisticated deep learning techniques. Similarity learning, a self-supervised machine learning, detects anomalous signals by estimating their similarity to background events. In this paper, we explore the potential of quantum computers for anomaly detection through similarity learning, leveraging the power of quantum computing to enhance the known similarity learning method. In the realm of noisy intermediate-scale quantum (NISQ) devices, we employ a hybrid classical-quantum network to search for heavy scalar resonances in the di-Higgs production channel. In the absence of quantum noise, the hybrid network demonstrates improvement over the known similarity learning method. Moreover, we employ a clustering algorithm to reduce measurement noise from limited shot counts, resulting in $9\%$ improvement in the hybrid network performance. Our analysis highlights the applicability of quantum algorithms for LHC data analysis, where improvements are anticipated with the advent of fault-tolerant quantum computers.  

 \end{abstract}
 
\maketitle
\tableofcontents

\section{Introduction}  

Since the discovery of the Higgs boson, the Large Hadron Collider (LHC) has been actively searching for signatures of beyond the Standard Model (BSM). These searches are typically based on prior knowledge of BSM properties. Despite extensive model-dependent searches, there has been no compelling evidence of new physics. 
One major challenge is the sheer number of possible new physics models, which makes it impossible to explore them 
comprehensively. 
Even if all theoretical hypotheses were tested, alternative scenarios might still be overlooked.

Anomaly detection offers a model-agnostic approach that does not require prior knowledge of the specific nature of BSM events. The general strategy in these analyses is to compare data directly with simulations across a large number of exclusive final states. Model-agnostic method has been performed by ATLAS \cite{ATLAS:2018zdn,ATLAS:2014sxa,ATLAS:2012qna} and CMS \cite{CMS:2017yoc,CMS:2011fra,CMS:2020ohc}. However, these approaches are sensitive to a large number of final states, which can lead to the look-elsewhere effect \cite{Gross:2010qma}, an increased probability of observing significant fluctuations simply due to the large number of compared distributions. Furthermore, these methods rely heavily on the accuracy of background simulations, and their validity depends on the accuracy of background modelling.

In practice, anomaly detection techniques at the LHC use advanced machine learning models that are trained on known backgrounds and then applied to detect events that deviate from the expected patterns \cite{CrispimRomao:2020ucc,Nachman:2020ccu,Belis:2023mqs,Belis:2023mqs,Kasieczka:2021xcg,Finke:2023ltw,Atkinson:2021nlt,Grosso:2023owo,Zipper:2023ybp,Roche:2023int,Chekanov:2023uot,Nachman:2020lpy,Golling:2023yjq,Bickendorf:2023nej,Finke:2023ltw,Krause:2023uww,Mikuni:2023tok,Buhmann:2023acn,Roy:2019jae,Knapp:2020dde}. 
These techniques include self-supervised learning methods, which do not require prior knowledge of the exact nature of the anomalies.  In particular, weakly supervised, density-based methods identify a signal phase space by comparing dense regions, which contain both signal and background, with low density regions containing background only. This method has been highly successful in applications such as signal bump hunting \cite{Metodiev:2017vrx,Collins:2018epr,Collins:2019jip,Andreassen:2020nkr,Hallin:2022eoq,Golling:2022nkl,Aguilar-Saavedra:2021utu}. The main advantage of this method is its independence from the accuracy of event simulations, allowing the model to be directly applied to the collected LHC data. However, this approach assumes that the signal is localized in a phase space, often identified using an invariant mass distribution, to pinpoint the signal region.

The major issue of the density-based approach is that the network performance depends on the phase space information. This means that a simple change in the four momenta of the final state particles changes the anomaly score. Similarity Learning (SL) methods mitigate this issue by identifying anomalous signal events based on their similarity to background events. This is achieved by employing a pair of neural network encoders that map input events into a latent space, ensuring that signal and background events are represented distinctly. 
SL methods have been used in LHC analysis in \cite{Dillon:2021gag,Esmail:2023axd} and for anomaly detection in \cite{Dillon:2023zac}.   

Recently, quantum computing has been proposed for anomaly detection in LHC analyses \cite{Ngairangbam:2021yma,Alvi:2022fkk,Araz:2022zxk,Schuhmacher:2023pro,Guan:2020bdl,Wu:2021xsj,Delgado:2022aty,Wozniak:2023xbe}.  
It has been demonstrated that quantum computers can effectively learn the similarity between different datasets \cite{Hammad:2023wme,Lloyd:2020eeh}. 
One of the most significant advantages of this approach is the potential for exponential speedup. Classical machine learning algorithms often scale inefficiently as dataset size or model complexity increases, whereas quantum algorithms can leverage superposition to perform multiple computations simultaneously. This could, in principle, lead to an exponential reduction in computation time for certain tasks. Additionally, quantum entanglement enhances the ability of quantum models to capture intricate correlations in the data, further boosting their performance in complex analyses.

Although classical SL learning and quantum methods have each demonstrated high performance,
a well-defined quantum SL approach for anomaly detection 
has yet to emerge. 
Employing quantum computing for SL has the potential to enhance the performance of each approach individually. On the one hand, the SL method is optimized to learn similarities between input events without labels; on the other hand, the quantum computer provides a high dimensional space, the Hilbert space of entangled qubits, to analyze the underlying structure of these events.

In this paper, we investigate the applicability of quantum computers for anomaly detection based on the SL method. 
Given the limitation of the current NISQ devices, we consider a hybrid classical-quantum network. Similar to the classical SL method, the hybrid network consists of a pair of Transformer encoders that map the input data into lower-dimension latent space. The mapped data is then encoded into the qubits with sequential unitary transformation. Applying a swap test, the output measurement quantifies the degree of similarity between the input events. This degree of similarity can be used as a cut-off score for detecting anomalous events. In fact, the measured similarity is significantly impacted by the noise in current quantum computers. This noise introduces substantial uncertainty in the similarity measurements, thereby reducing the network's efficiency. We propose a classical clustering method to alleviate the uncertainty caused by the currently unavoidable quantum noise.

This paper is structured as follows. In section \ref{sec:2} we discuss the SL methods elaborating the anomaly detection using classical and quantum SL approaches. In section \ref{sec:3}, we outline the strategy for the numerical analysis.  Specifically, we consider the process of heavy scalar resonance decaying to di-Higgs at the LHC. 
The results are given in section \ref{sec:4} and the conclusion in section \ref{sec:5}.

\section{Similarity learning}
\label{sec:2}
SL is designed to learn effective data representations by comparing pairs of similar and dissimilar features. To achieve this, an SL network employs two encoders that process input data in pairs. By contrasting positive pairs, i.e.\ those with similar information, with negative pairs with dissimilar information, the network learns to distinguish between signal and background events without requiring labels. This is achieved by mapping input pairs into a latent space, where positive pairs are positioned close together, while negative pairs are pushed farther apart.

The construction of positive and negative pairs is critical for the success of SL tasks and must be carefully addressed. Since no labels are used in this approach, we specifically assume that positive pairs consist of matching each event in the dataset with an augmented version of itself, while negative pairs are created by pairing each event with all other events, excluding itself or its augmented versions.
The augmentations provide various perspective and transformations of the same event, enabling the network to learn robust and meaningful representations of unlabeled data. Importantly, these data augmentations should be Lorentz invariant to ensure consistency with physical principles. We consider three augmentation functions as in \cite{Dillon:2023zac}:
\begin{itemize}
    \item Randomly rotate the azimuthal angle $\phi$ of each particle in the event with angle sampled from $[0,2\pi]$.
    \item Smearing the $\eta$-$\phi$ coordinates ($\eta$ being the pseudo-rapidity) of each particle in the event by resampling from a Normal distribution centred on the original $\eta$-$\phi$ values, with a variance equal to the inverse of the particle's transverse momentum.
    \item Smearing the transverse momentum $p_T$ of each object by resampling from a Normal distribution centred on the original value,
    with a variance of the inverse of the transverse momentum of the particle.
\end{itemize}

These augmentations reflect the imperfection of the detectors and the symmetry of the events under the rotation of the azimuth angle.
Even though the augmented version of the event may appear different from the original event, it still stems from the same interaction process and should be recognized as similar by the network. 

Once positive and negative pairs are projected onto the latent space of the network via encoders,  minimizing (maximizing) the distance in the network latent space between the positive (negative) pairs can be done using a contrastive loss function. Alternatively, one can
embed the projected latent space data onto a Variational Quantum Circuit (VQC) and minimize a Hilbert-Schmidt (HS) loss function, which is a metric function that measures the distance between the input pair. 
This approach can be considered as the quantum version of classical SL learning. 

In this section, we describe the anomaly detection method using SL techniques. We consider two networks, a classical SL network proposed in \cite{khosla2020supervised} and a hybrid classical-quantum network proposed in the current paper. In both approaches a pair of Transformer encoders with shared weights are used to encode the high-dimensional input data into a latent space of the network. 

\subsection{Classical similarity learning}%

At the heart of SL lies the idea that representations of augmented versions of the same instance (positive pairs) should be pulled closer in the latent space, while representations of different instances (negative pairs) should be pushed apart. The model is optimized by contrasting these examples via minimizing a contrastive loss function. This loss function operates in the latent space where data points are projected by the network encoders.

During each training iteration, we apply random augmentations to the input data. This results in two distinct but physically equivalent views for each event in the training batch.  The event pair, original and augmented,  passes through the two encoder networks, generating two feature representations. Since the encoder weights are shared between the event pair, the model learns invariant representations robust to the applied physical augmentations. The feature representations are then passed through a projection head, a fully connected layer, which projects the representations into a new space where the contrastive loss function will be applied. 

Once the embeddings for all events in the training batch are computed, the contrastive loss is applied to maximize the similarity between positive pairs and minimize the similarity between negative ones. Contrastive loss has the form 
\begin{equation}
  \mathcal{L}_{\rm contras}  = -\sum\limits_{i \in N}\log  \frac{\exp \left(s(z_i, z_i^\prime)/\tau \right)}{\sum\limits_{i\neq j \in N}  \exp \left(s(z_i, z_j)/\tau \right)}\,,
  \label{eq:1}
\end{equation}
where the primed element indicates the augmented version of the event and $\tau$ is a regularization parameter. The similarity function $s(\cdot, \cdot)$ is the cosine similarity between the input pairs
\begin{equation}
    s(z_i,z_j) = \frac{z_i\cdot z_j}{\lVert z_i \rVert\lVert z_j \rVert } = \cos\phi_{ij}\,,
\end{equation}
with $\phi_{ij}$ defining the similarity magnitude between the two events, i.e., $\phi_{ij}\sim 0$ when the two events are similar, while $\phi_{ij}\sim 1$ when the two events originate from different classes. 
The regularization parameter $\tau$ controls the sharpness of the similarity distribution: a smaller $\tau$ makes the similarity distribution sharper, placing more emphasis on ``hard'' negative events, those that are somewhat similar but not identical. It balances the trade-off between focusing too much on negatives that are far away or over-emphasizing hard negatives, which can be difficult to separate.

After the network is trained, the projection head is discarded, and the encoder network is used to embed the inputs into a latent space.  The embedded data are evaluated by freezing the encoder\footnote{This means that the weights of the encoder are no longer updated in the subsequent steps.} and training a simple Linear Classifier (LC) on top of the learned embeddings.  A simple LC consists of a single fully connected layer with softmax activation for classification tasks; this takes the feature vectors produced by the frozen encoder and tries to map them to the desired output classes, signal or background based on pseudo labeled data.

This is a common evaluation technique called linear evaluation. The idea here is to assess the representations learned by this encoder without further modifying them. The linear evaluation protocol is based on the assumption that a well-trained encoder will produce high-quality, discriminative features, even for complex datasets. These features should ideally contain enough information such that a simple linear model can easily separate different classes. If the learned features are good, the LC will perform well, even though it has only a linear layer without any hidden layers and has limited complexity. 

\subsection{Quantum similarity learning}

\begin{figure*}[t!]
    \includegraphics[width=\textwidth]{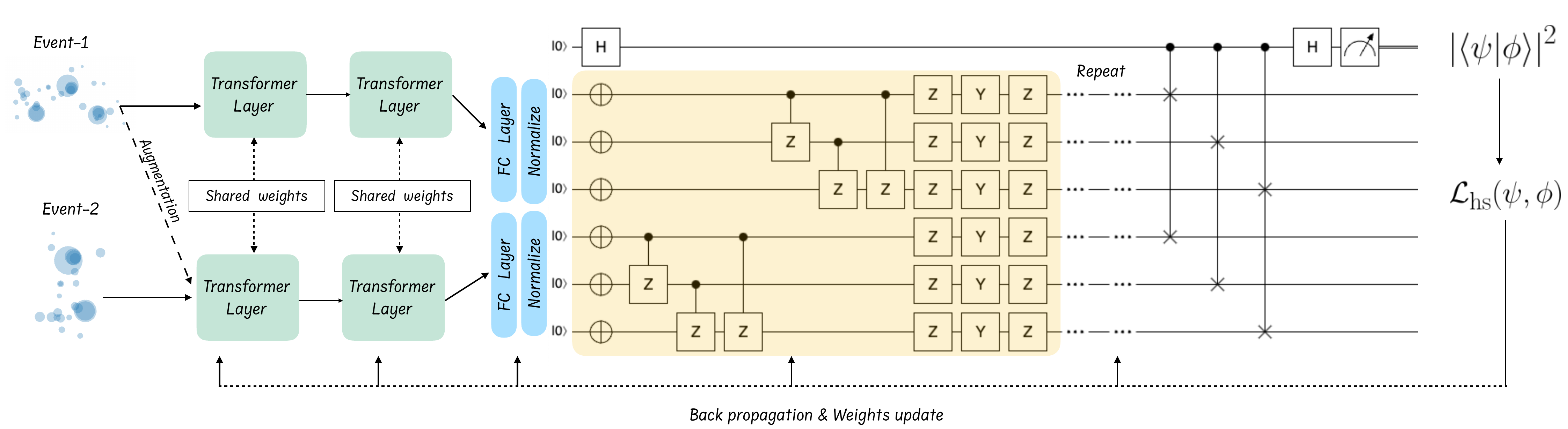}
    \caption{Schematic architecture of the hybrid classical-quantum network for a VQC with $7$ qubits. The input to the network consists of pairs of events: a positive pair, which consists of the original event and its augmented version, and a negative pair, which consists of the original event with another event in the training batch. Two Transformer encoders with shared weights are highlighted in green. 
    A projection normalization layer applied to the Transformer outputs is highlighted in blue. The VQC highlighted in yellow uses $6$ registered qubits and an ancillary qubit for measurement. Filled circles represent strong entanglement with CZ gates, and  CSWAP gates are added at the end of the VQC.  The network's output is the fidelity between the input events, which quantifies the similarity between the two inputs. Visualization of the VQC is done by using a quantum circuit simulator \href{https://algassert.com/quirk}{Quirk}. }
    \label{fig:1}
\end{figure*}

Quantum computers offer an alternative approach for anomaly detection by constructing a hybrid classical-quantum network for SL, as shown in Figure \ref{fig:1}. 
Similar to the classical SL, the hybrid version works by using a pair of classical encoders with shared weights. The encoders map the input data into a lower-dimensional space. The dimension of the latent space is fixed by the number of the used qubits. A VQC with one ancillary qubit is then used to measure the similarity between the embedded data by the classical encoders.  The VQC first encodes the mapped data onto the designated qubits. Once the data is encoded, parametric unitary gates are applied to process it. To capture the complexity of the input data, the VQC can be repeated multiple times, with the number of repetitions to be optimized for the best performance. Finally, to measure the similarity between the input pair, a controlled-swap (CSWAP) test circuit is used \cite{Fredkin2002ConservativeL}. The CSWAP test calculates the overlap,  quantum fidelity, between the quantum states by projecting the states onto an ancillary qubit. The final output measures the similarity between the two states with probability $\mathcal{P}=|\langle \psi|\phi\rangle|^2$, where  $\mathcal{P} =1$ indicates the two states are identical and  $\mathcal{P}=0$ indicates completely different states.  During the training, a classical optimizer is used to optimize the free parameters of the classical encoders and those of the VQC.\footnote{VQC also called parameterized quantum circuit in which the unitary gates are parameterized by free parameters, $\theta$, that control the rotation magnitude of the qubit.} As a result, the hybrid network takes pairs of positive or negative events and tries to maximize (minimize) the quantum fidelity between positive (negative) pairs. 

The key point of VQC training lies in the initial encoding of the classical data. 
This encoding can be viewed as a mapping of the classical data into the Hilbert space of the qubits. In this context, data encoding is not merely a preparatory step; it acts as a feature map that alters the structure of the classical data in a non-trivial way. Specifically, the encoding gates apply nonlinear operations to the data, which can alter the distances between the classical data points once mapped onto the qubit Hilbert space. 
If this mapping makes the data linearly separable, it enhances the expressive power of the VQC and improves its classification capabilities. Traditional data encoding does not change the distance between the states from the different classes, because it utilizes unitary quantum gates acting on quantum states, and hence linear operations, 
see Appendix \ref{App:A} for more details. 
However, VQC with ``variational encoding'' and repeated data embedding offers a novel approach by training the mapping process to maximize the distance between data points from distinct classes, thereby improving classification performance.  Variational encoding  can be written as  a sum of a multi-dimensional partial Fourier series \cite{schuld2021effect}
\begin{equation}
    f_\theta(x) =\sum_{\omega=-L}^L c_\omega(\theta)e^{i\omega x} \,,
\end{equation}
with $L$ is the number of the repeated blocks. The frequency spectrum is determined by the eigenvalues of the data encoding gates, while the Fourier coefficients depend on the overall circuit structure. Since quantum models yield real-valued outputs, the learned function can be expressed as a linear combination of sine and cosine terms, $\cos(\omega\,  x_i)$ and $\sin(\omega\, x_i)$.  
The representation of quantum models as Fourier sums offers valuable insights into the function space the model can learn. 
The frequency spectrum defines the exponential part of the accessible functions in the quantum circuit, while the Fourier coefficients dictate how these functions are combined. This perspective reveals the function classes that the quantum model can learn. 
With repeated encoding layers, the VQC behaves like a Fourier sum of trigonometric functions, where the embedding blocks determine the frequencies, and the variational blocks govern the amplitude. 
Repeating this process increases the nonlinearity of the learned function, boosting the expressiveness needed for high-performing classification.

The structure of a single VQC comprises three main parts: a variational data encoding layer $U(x,\theta)$, a strong entangled layer and a variational layer $U(\phi)$. 
Each of the three components, highlighted in yellow in Figure \ref{fig:1}, will be described in the following.

{\textbf{Variational encoding layer}} in which both the data encoding process and the trainable parameters are intertwined, allowing the circuit to optimally learn how to encode the classical data. Encoding process acting on registered qubits in an initial state  $|0\rangle^{\otimes n}$ can be expressed as \cite{perez2020data}
\begin{equation}
     \begin{split}
    &|\psi_{\rm encoded}(x,\theta)\rangle = U(x,\theta)|0\rangle^{\otimes n} = \\
    &\left( \prod\limits_{i,j,k=1}^n R_Z(x_i+\theta_i)R_Y(x_j+\theta_j)R_Z(x_k+\theta_k)\right) |0\rangle^{\otimes n}\,,
    \end{split}
\end{equation}
where $n$ represents the number of the qubits and $R_Z$,$R_Y$ are rotational gates around the Z- and Y-axis of the Bloch sphere, respectively; $x$ represents the latent space data that are mapped by the  encoders pair, and $\theta$
are the trainable parameters that optimize the data encoding during the training process.

{\textbf{Strong entangled layer}} increases the expressive power of a quantum circuit by increasing the number of the entangled qubits $n$ in which the resulting states are $2^n$. 
Entanglement can enhance the expressive power of the circuit because the system can explore more complex correlations between qubits, which is necessary for representing complex quantum states. For this purpose, we use the non-parametric Controlled-Z (CZ) gate, a two-qubit gate that applies the Pauli-Z rotation to the target qubit. A sequence of CZ gates is applied to all neighbouring qubits in the VQC, resulting in a transformation of the encoded quantum state  as 
\begin{equation}
    |\psi_{\rm entangled}\rangle = \prod\limits_{i=1}^{n-1} CZ_{i,i+1} \left(\bigotimes_{j=1}^n|\psi_{\rm encoded} \rangle_j \right)\,.
\end{equation}

{\textbf{Variational layer}} is added to the quantum circuit after the entanglement. This layer consists of parameterized quantum gates that are optimized during training. The final quantum states can be expressed as 
\begin{equation}
    |\psi_{\rm final}\rangle = U(\phi)|\psi_{\rm entangled}\rangle\,,
\end{equation}
with $U(\phi)$ being a unitary rotation defined in Appendix \ref{App:A}.

The VQC can be repeated multiple times to improve the expressive power of the network. Finally, to estimate an overlap between the processed quantum states $|\psi^1_{\rm final}\rangle$ and $|\psi^2_{\rm final}\rangle$,\footnote{These states represent the event pair, either positive or negative pair, being processed by the classical  encoders pair and the repetitions of VQCs.} we use a CSWAP test with an ancillary qubit for measurement \cite{Buhrman:2001rma}. A measurement on this ancillary qubit in a computational basis provides an estimation of fidelity between two states as $|\langle \psi^2_{\rm final}| \psi^1_{\rm final}\rangle|^2$. In this case, the network tries to maximize (minimize) the fidelity between the two quantum states if they belong to a positive (negative) pair. As the calculation for the fidelity is based on the probability, one needs to perform multiple CSWAP tests, and the uncertainty in the fidelity decreases as $O(1/\sqrt{k})$ with $k$ repetitions of the CSWAP test. 

The output of the hybrid classical-quantum network, with a VQC of $7$ qubits, can be expressed as
\begin{equation}
   \begin{split}
  \langle \psi_{\rm final}|\sigma_z | \psi_{\rm final}\rangle& \hspace{4mm} \rm with \\
  &| \psi_{\rm final}\rangle= \prod_i ^{m-3} \rm CWAP |c, \psi^1_i,\psi^2_{i+3}\rangle\,, \end{split} 
\end{equation}
where $m$ is the number of the registered qubits,  $c$ is the ancillary qubit and the quantum states, and $\psi^1,\psi^2$ are expressed as 
\begin{equation}
     |\psi^{1,2}\rangle = U(\phi)\left( \prod\limits_{i=1}^{n-1} CZ_{i,i+1}\  U(\mathcal{T}(x^{1,2}),\theta)|0\rangle^{\otimes n} \right)\,,
\end{equation}
where $n$ is the number of the registered qubits, $\mathcal{T}$ represents the classical Transformer encoder as described in Appendix \ref{subsec:1.3}, and $x^{1,2}$ is the input event pair, either positive or negative. The output probability measures the similarity between the input event pair, with $\mathcal{P} = 1$ when the two events are identical and $\mathcal{P} = 0$ when the two events are different.   

A training metric loss function is used to maximize the similarity between positive pairs during the training process. The purpose of this loss function is to map events from positive pairs close together while pushing apart the negative pairs on the Hilbert space of the qubit.  In supervised learning, where labels are provided to identify the signal and background events during training, a loss function based on Hilbert-Schmidt (HS) distance can be used. The HS distance is given by
\begin{equation}
    \mathcal{D}_{\rm HS}(\rho,\sigma) = \text{Tr}\left[ (\rho-\sigma)^2 \right]\,,
    \label{eq:8}
\end{equation}
where $\rho$ and $\sigma$  are the mixed density matrices of each event in the pair in the training batch:
\begin{equation}
    \rho = \frac{1}{M}\sum\limits_{i\in \rm batch}| \psi^1\rangle\langle \psi^1| \hspace{3mm} \text{and} \hspace{3mm} \sigma = \frac{1}{M}\sum\limits_{j\neq i\in \rm batch}|\psi^2\rangle\langle \psi^2|\,,
\end{equation}
with $i,j$ running over the batch size $M$.
Recalling that the quantum fidelity between  pure  states can be expressed as $|\langle \psi^1|\psi^2\rangle|^2 = \rm Tr (\rho\sigma)$ \cite{kobayashi2003quantum}, HS loss function can be defined as \cite{Wang:2021shr} 
\begin{equation}
   \begin{split}
    \mathcal{L}_{\rm HS}^{\rm supervised} &= 1- \frac{1}{2}\mathcal{D}_{\rm HS}\\ & = 1- \frac{1}{2} \left[ \rm Tr(\rho^2)+ \rm Tr (\sigma^2)\right] + \rm Tr (\rho\sigma)\,,\end{split} 
\end{equation}
with minimum value of the loss function at $\rm Tr(\rho^2)\sim 1$, $\rm Tr(\sigma^2) \sim 1$ and $\rm Tr (\rho\sigma) \sim 0$. In this case, the supervised network aims to maximize the purity of the measured signal and background events while minimizing their overlap by mapping them to distinct regions within the Hilbert space of the qubit. For anomaly detection, with no labels provided, we use a modified loss function
\begin{eqnarray}
    \mathcal{L}_{\rm HS}(\rho,\sigma) =  1- \text{Tr}(\rho\rho^\prime)  + \text{Tr}(\rho\sigma)\,,
    \label{eq:11}
\end{eqnarray}
where $\rho^\prime$ represents the augmented event  of $\rho$ from a positive pair. The loss function has a minimum value if the measured quantum fidelities are $\text{Tr}(\rho\rho^\prime)=1$ and $\text{Tr}(\rho\sigma)=0$. In this case, the network maps events, and their augmented versions, from different classes far apart from each other on the Hilbert space of the ancillary qubit. Notably, we omitted the term $\rm Tr(\sigma^2)$ in the supervised loss, as the SL task involves only positive pairs, represented by $\rm Tr(\rho\rho^\prime)$, and negative pairs, represented by $\rm Tr(\rho\sigma)$.

After training, events from different classes, signal and background, are clustered away from each other on the Hilbert space of the qubit, as shown in the middle plot of Figure \ref{fig:5}. The two clusters are then used to test the network performance by computing how close the test event is to each of the two clusters. 
This can be achieved by a fidelity classifier test \cite{Wang:2021shr}. A fidelity classifier is defined as the difference  between squares of the inner product between the embedded test sample $|x\rangle$ and the respective class encoded by the training samples, $| \psi^1_{\rm trained}\rangle$ and 
 $| \psi^2_{\rm trained}\rangle$, as \cite{Buhrman:2001rma,Lloyd:2020eeh}
\begin{equation}
    \mathcal{F}(x) = \frac{1}{M}\sum \left|\langle x| \psi^1_{\rm trained}\rangle \right|^2 - \frac{1}{M}\sum \left|\langle x| \psi^2_{\rm trained}\rangle \right|^2\,.
    \label{eq:fid}
\end{equation}

For binary classification, the fidelity classifier assigns a binary predicted label to the input data according to 
\begin{equation}
\hat{Y}  = 
\begin{cases}
            -1 \text{\  if \ } \mathcal{F}(x) < 0 \\
              +1 \text{\  if \ } \mathcal{F}(x) \ge 0
    \end{cases} .
\end{equation}
The output ranges between $[-1,1]$ with $\mathcal{F}(x) < 0$ ($\mathcal{F}(x) > 0$), indicating the test event most likely belongs to the first (second) class.

\section{Analysis setup}%
\label{sec:3}
In this section, we outline the strategy for our numerical analysis. We focus on the self-supervised study of events with a final state of four leptons and two $b$-jets at the High-Luminosity (HL)-LHC. The anomalous signal arises from the decay of a heavy scalar resonance into a pair of Standard Model-like Higgs bosons, through the process $g g \to H \to h_{\rm SM} h_{\rm SM}$, with $h_{\rm SM} \to Z^\ast Z \to 4\ell$ and $h_{\rm SM} \to \bar{b}b$. We adopt signal benchmark points from the Two Higgs Doublet Model (THDM) with Type-II Yukawa coupling. While the di-Higgs final state is a very difficult final state, it has received attention recently in relation to the Higgs self-coupling, which has not been explored yet. 

We employ two approaches for anomaly detection analysis: a classical network and a hybrid classical-quantum network. Both utilize Transformer encoders to map high dimensional input data into a lower dimensional latent space. The hybrid model combines a classical Transformer encoder with VQCs. To demonstrate the advantages of deep VQCs with a larger number of qubits, we explore hybrid networks with VQCs consisting of 7 and 11 qubits. In both configurations, the first qubit is an ancillary qubit for measurement, while the remaining qubits serve as register qubits.

\subsection{Data simulation}%
For event simulations, the THDM Lagrangian is implemented into SARAH \cite{Staub:2008uz} to produce all coupling of interaction vertices of the model whose numerical values are computed with SPheno package \cite{Porod:2011nf}. 
MadGraph5 \cite{Alwall:2011uj} is used to calculate the cross section and parton level events, and PYTHIA \cite{Sjostrand:2007gs}  is used for parton shower, hadronization, heavy flavour decays, and adding the soft underlying event. 
Jets are reconstructed from particle flow objects simulated by DELPHES \cite{deFavereau:2013fsa} detector simulation package. We use FastJet \cite{Cacciari:2011ma} package for jet reconstruction with anti-$K_T$ clustering algorithm of $R=0.4$ \cite{Cacciari:2008gp}. 

We use TensorFlow framework \cite{tensorflow2015-whitepaper}  to construct the classical SL network.  Scikit-Learn \cite{Pedregosa2011ScikitlearnML} is used to 
facilitate network processing and evaluation. 
We use PennyLane \cite{Bergholm:2018cyq} framework for VQC implementation. 
We use the TensorFlow interface in PennyLane 
to construct and train the hybrid classical-quantum SL network.\footnote{Constructing the VQC as TensorFlow layer, PennyLane requires Keras version 2. In this work, we use TensorFlow version 1.14.0 in which Keras 2 is part of it. } 

\subsection{Heavy scalar search}

We align our analysis with the latest ATLAS/CMS heavy resonance search results \cite{ATLAS:2024bzr,CMS:2022omp}. The main background source arises from the production of a single SM Higgs boson decaying to $ZZ^\ast$, with two jets originating from QCD radiations. Another significant contribution is the continuum background process $pp \to ZZ^\ast$. Irreducible backgrounds such as $t\bar{t}Z$, $t\bar{t}W$, and three-gauge boson production are also included in the analysis. Although these irreducible backgrounds can produce the same final state as the signal, they have much smaller cross sections and are thus subdominant.

Anomalous signal events are considered from a heavy scalar resonance in the THDM with Type-II Yukawa couplings. To ensure that the analysis is independent of the heavy scalar mass, we examine three masses: $m_H = 0.6$, $0.8$, and $1$ TeV. These three signal samples are combined into a single dataset representing the anomalous signal. The chosen benchmark points satisfy all theoretical and experimental constraints, as adopted from \cite{Hammad:2022wpq, Hammad:2023sbd}.

Selected events are required to contain at least four isolated leptons, electrons or muons, with opposite charges, as well as two jets. To ensure accurate jet measurements and minimize pileup effects, all jets must have a transverse momentum $p_T \geq 20$ GeV. Only leptons with $p_T \geq 7$ GeV are considered, with the additional requirement that the leading lepton must have $p_T \geq 20$ GeV and the second leading lepton must have $p_T \geq 15$ GeV. All leptons must be separated by $\Delta R (l_i, l_j) \geq 0.02$, and electrons and muons must be separated by $\Delta R(e, \mu) \geq 0.05$, where the angular distance is defined as $\Delta R \equiv \sqrt{(\Delta \eta)^2+ (\Delta \phi)^2}$.

In each event, all permutations of the selected leptons are considered for reconstructing the $Z$ boson pair. For each pair, the lepton combination with an invariant mass closest to the $Z$ boson rest mass is used to reconstruct the on-shell $Z$ boson, with a lower mass cut of 40 GeV. The remaining lepton pair is assigned to reconstruct the off-shell $Z^\ast$ boson, with a lower mass cut of 12 GeV. If more than one $ZZ^\ast$ candidate passes the selection criteria, the one closest to the SM Higgs boson mass, 125 GeV, is chosen. To suppress QCD background from soft leptons originating from hadron decays, all considered lepton pairs must have an invariant mass greater than 4 GeV. Additionally, to enhance di-Higgs reconstruction, the distance between leptons and jet candidates must satisfy $\Delta R(l,j) \geq 0.3$.

Events that pass the selection criteria are preprocessed to match the input dimensions of the networks being considered. Since both classical and quantum networks utilize the same input, accessed via Transformer encoders, a single dataset is used for both networks. The input dataset is structured with dimensions $d = (n,p,f)$, where $n$ is the number of events, $p$ represents the dimension of the particle tokens, and $f$ represents the dimension of feature tokens assigned to each particle. We consider particle tokens with dimension $p = 11$: four leptons, two jets,  $Z$ boson pair,  $h_{\rm SM}$ pair and the heavy scalar. All bosons are reconstructed from the four-momenta of their final-state particles. Each particle token is assigned six features: transverse momentum, pseudorapidity, azimuthal angle, energy, mass, charge, and particle ID. After the data is prepared, we combine signal and background events in one data set and shuffle them. As a self-supervised task, the combined dataset has no labels to identify the different events.

\subsection{Network  architecture and training}%

Both classical and quantum networks use the same Transformer encoders to map the high-dimensional input data into the latent space of the network. The Transformer encoder consists of an input normalization layer followed by two sequentially repeated Transformer layers. Each Transformer layer includes a multi-head attention mechanism with eight attention heads. The output from the multi-head attention is added to the original input via a skip connection, ensuring dimensional consistency since the output of the attention heads matches the dimensions of the input dataset, see subsection \ref{subsec:1.3} for details.
The resulting output is then passed through a Multi-Layer Perceptron (MLP) with two fully connected hidden layers containing $128$ and $6$ neurons, respectively, and activated by the ReLU function. A second skip connection adds the MLP output to the output of the multi-head attention. Notably, the output of each Transformer layer retains the same dimensions as those of the input, allowing for the repetition of the Transformer layers.
The final output of the stacked Transformer layers is passed through a normalization layer, which normalizes the $L_2$  norm to one. This normalization is crucial for computing similarity in the classical contrastive loss or for further mapping to the VQC. 

\begin{figure}[th!]
    \includegraphics[width=0.45\textwidth]{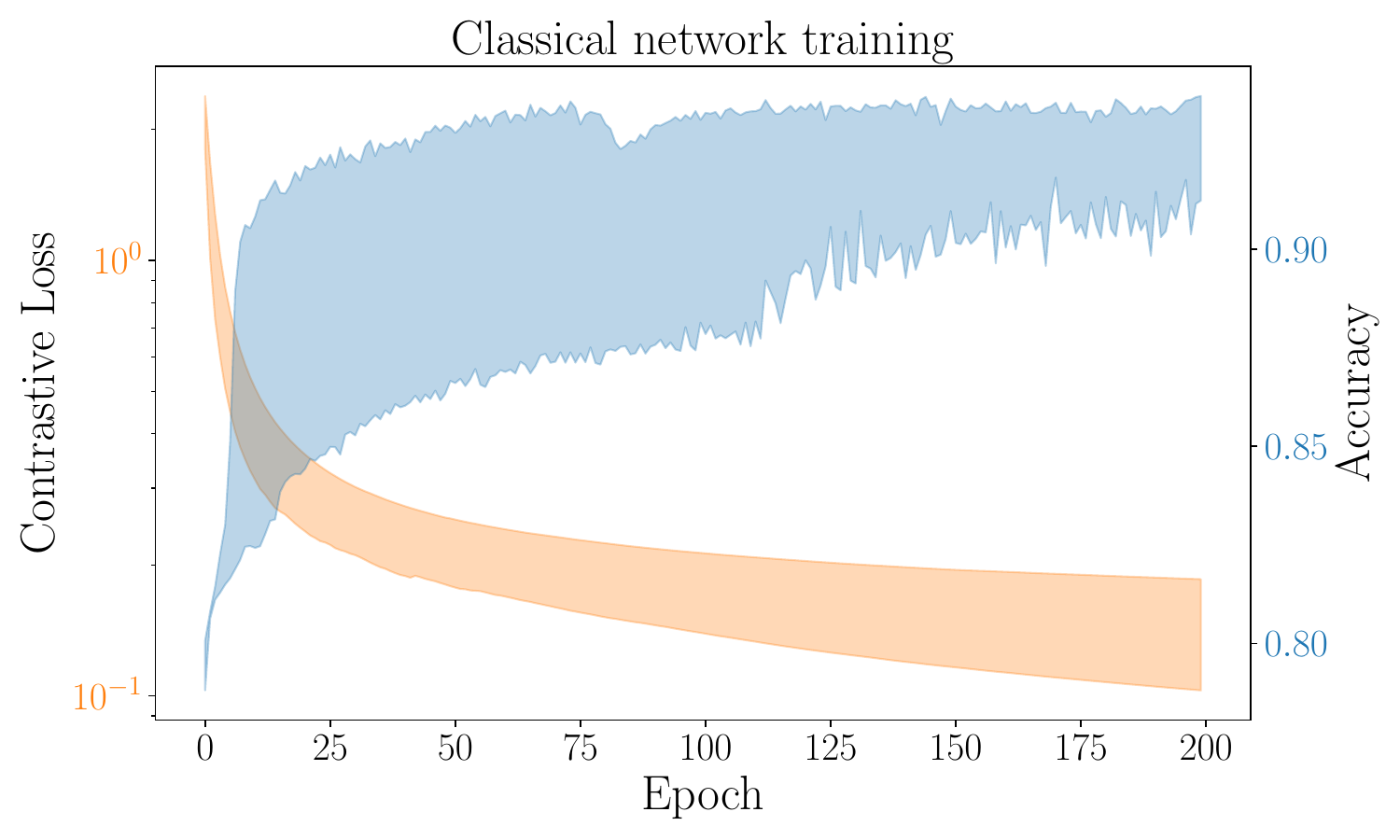}
    \caption{Training metrics of the classical SL network as a function of training epochs. The orange distribution represents the contrastive loss at each epoch, with values indicated on the left vertical axis. The blue distribution represents the accuracy of the linear classifier, with values shown on the right vertical axis. The shaded bands represent the range between the minimum and maximum values across three runs with different random seeds.}
    \label{fig:2}
\end{figure}
For the classical SL network, a linear projection head with $500$ neurons is added. The output of the projection head from each of the Transformer encoders is used to compute the contrastive loss function, Eq.~\eqref{eq:1}, during the training process.   

For the hybrid classical-quantum network with $7$ $(11)$ qubits, a linear projection head with $9$ $(15)$ neurons is added. The output of the projection head from the two Transformer encoders is concatenated in one vector and mapped to the VQC. The VQC is repeated $4$ times before the CSWAP gate for fidelity measurement. The final output is then used to minimize the HS loss function
defined in Eq.~\eqref{eq:11}.

\begin{figure}[th!]
    \includegraphics[width=0.45\textwidth]{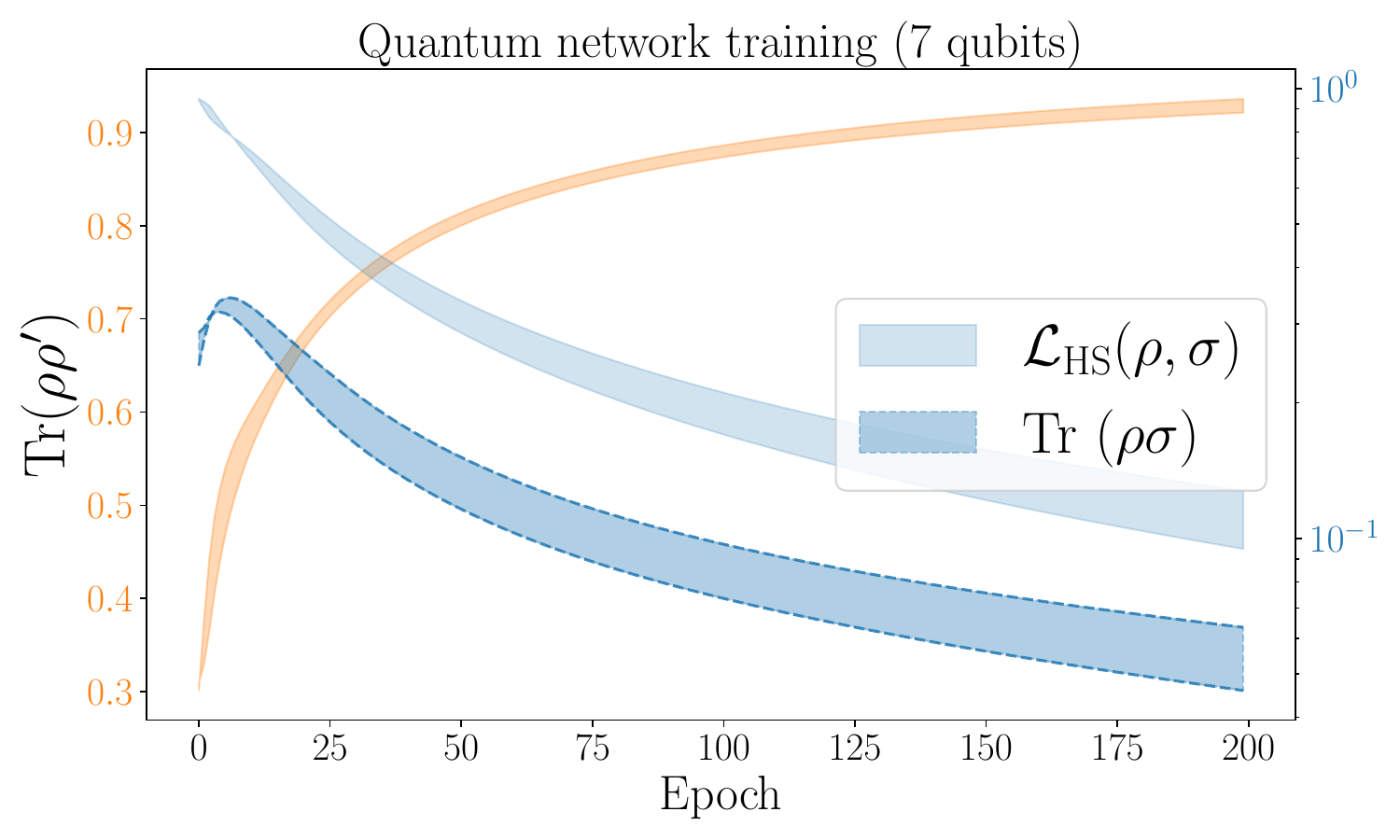}\\
    \includegraphics[width=0.45\textwidth]{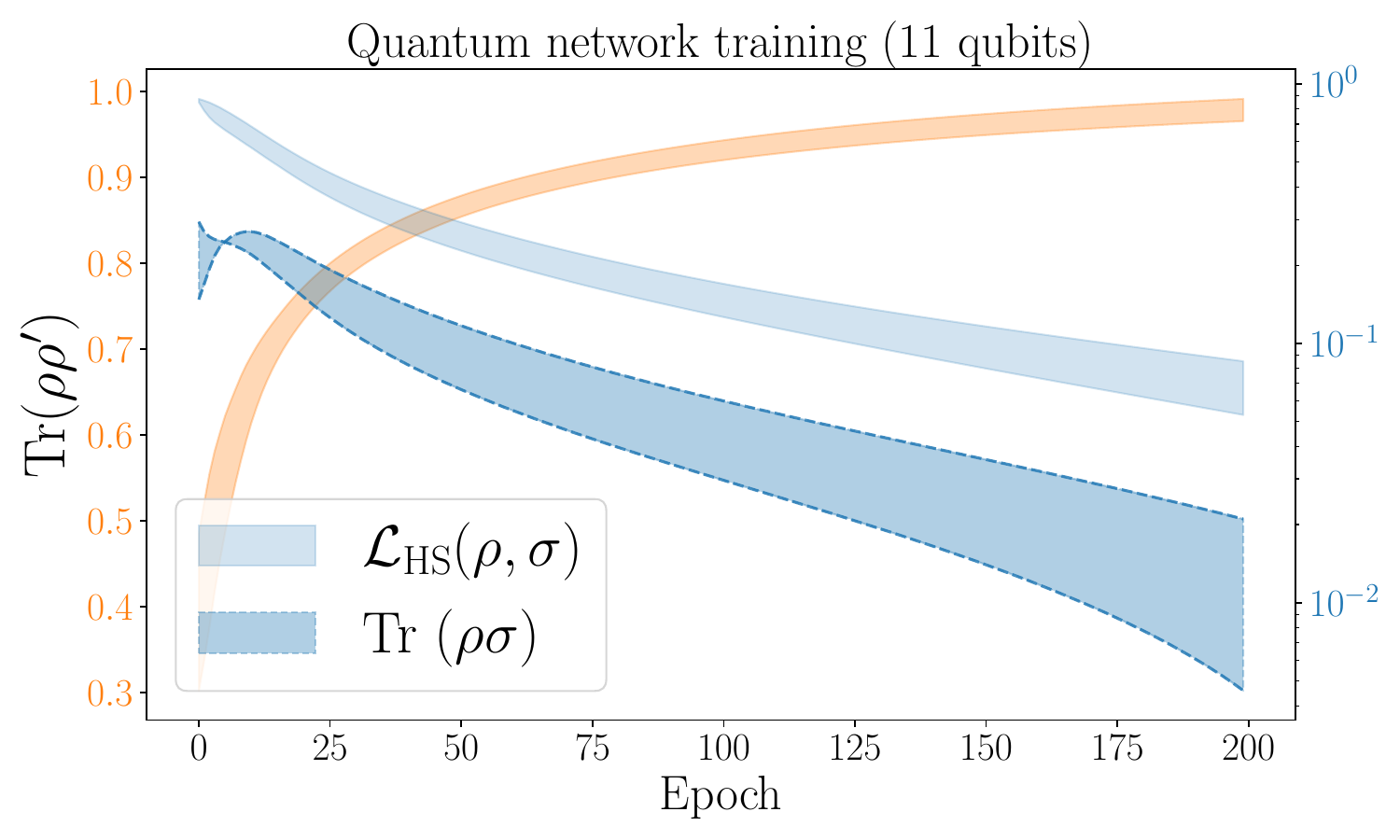}
    \caption{Training metrics of the quantum SL network are shown as a function of the number of training epochs for a smaller VQC with $7$ qubits (top plot) and a VQC with $11$ qubits (bottom plot). The light blue distribution represents the purity of the measured quantum states for positive pairs, $\rm Tr(\rho\rho^\prime)$. The dark blue distribution, with dashed edges, shows the overlap between the measured quantum states of negative pairs, $\rm Tr(\rho\sigma)$. The orange distribution represents the total loss function. 
    The left vertical axis shows the values for the orange distribution, while the right vertical axis shows the values for the blue distribution. 
    The bandwidth in each distribution indicates the range between the minimum and maximum values across three runs with different random seeds.}
    \label{fig:3}
\end{figure}

For both network training sessions, a dataset of $200,000$ events is used, with the Adam optimizer \cite{kingma2014adam} employed to minimize the respective loss function. The networks are trained for $200$ epochs with a batch size of $128$. Figure \ref{fig:2} illustrates the minimization of the classical contrastive loss function during training, shown in orange on the left vertical axis. The accuracy of the LC used to validate the network, is displayed in blue on the right vertical axis. The shaded regions in both distributions represent the range of minimum and maximum values across three training runs with different random seeds. The accuracy of the LC shows significant variation due to its simplistic design; it consists of a single fully-connected linear layer without non-linear parameters, limiting its ability to capture the complex structure of the input data. However, the key observation is that, by the end of training, all runs converge to high accuracy.  

The training metrics for the hybrid classical-quantum network are shown in Figure \ref{fig:3} for a VQC with 7 qubits (top plot) and a VQC with 11 qubits (bottom plot). In both plots, the dark blue distribution represents the overlap between negative pairs, which decreases during training. In contrast, the purity of the positive pairs increases. This indicates that during training, the network enhances the purity of positive pairs while minimizing the overlap between negative pairs, effectively mapping similar events closer together and pushing dissimilar events farther apart in the Hilbert space. The total HS loss function, shown in light blue, exhibits a continuous decrease throughout the training process.

A more complex VQC with a larger number of qubits can represent more complicated data compared to one with fewer qubits. This arises from the increased dimensionality of the Hilbert space and the greater expressibility of quantum states within that space. A larger number of qubits allows for more entanglement and provides additional degrees of freedom for the unitary transformation $U(\theta)$, which enhances the VQC's ability to represent complex data. Specifically, as the number of qubits increases, so does the number of parameters $\theta$, enabling more intricate transformations of the input data. This explains the improvement in network performance as the number of qubits in the VQC increases.

\section{Results}
\label{sec:4}

In this section, we test the performance of different networks when applied for the search for anomalous data in the four-leptons and two-jets final state. 
The discriminative power of each network will be a measure of how well the signal and background can be characterized through their different features, all entangled together into several low-level information of the final state particles in each event. 
This can be quantified by computing the Receiver Operating Characteristic (ROC) curve. The ROC curve visualizes the True Positive Rate (TPR), signal events that are correctly identified by the network, against the False Positive Rate (FPR), the signal events that are incorrectly identified by the network as background events, for different threshold values of the network output. 
The better the discrimination performance between signal and background events, the higher the TPR over the FPR.  

For the classical SL network, after the network training, LC layer is added to one of the Transformer encoders.\footnote{As both Transformer encoders share their weights, it does not matter which Transformer encoder we choose.} This new setup is used to test the network performance and is used for anomaly detection tasks. 

For the quantum network, the trained dataset is separated into two groups according to the fidelity measurements. This can be done by matching the trained events with $\mathcal{F} > 0.5$, which indicates the events from the same class, to their counterpart with $\mathcal{F} < 0.5$. In this case, training events are divided into two classes, signal and background, in which we can use the fidelity classifier to evaluate the network performance as mentioned in Eq.~\eqref{eq:fid}. 

\begin{figure}[th!]
    \includegraphics[width=0.45\textwidth]{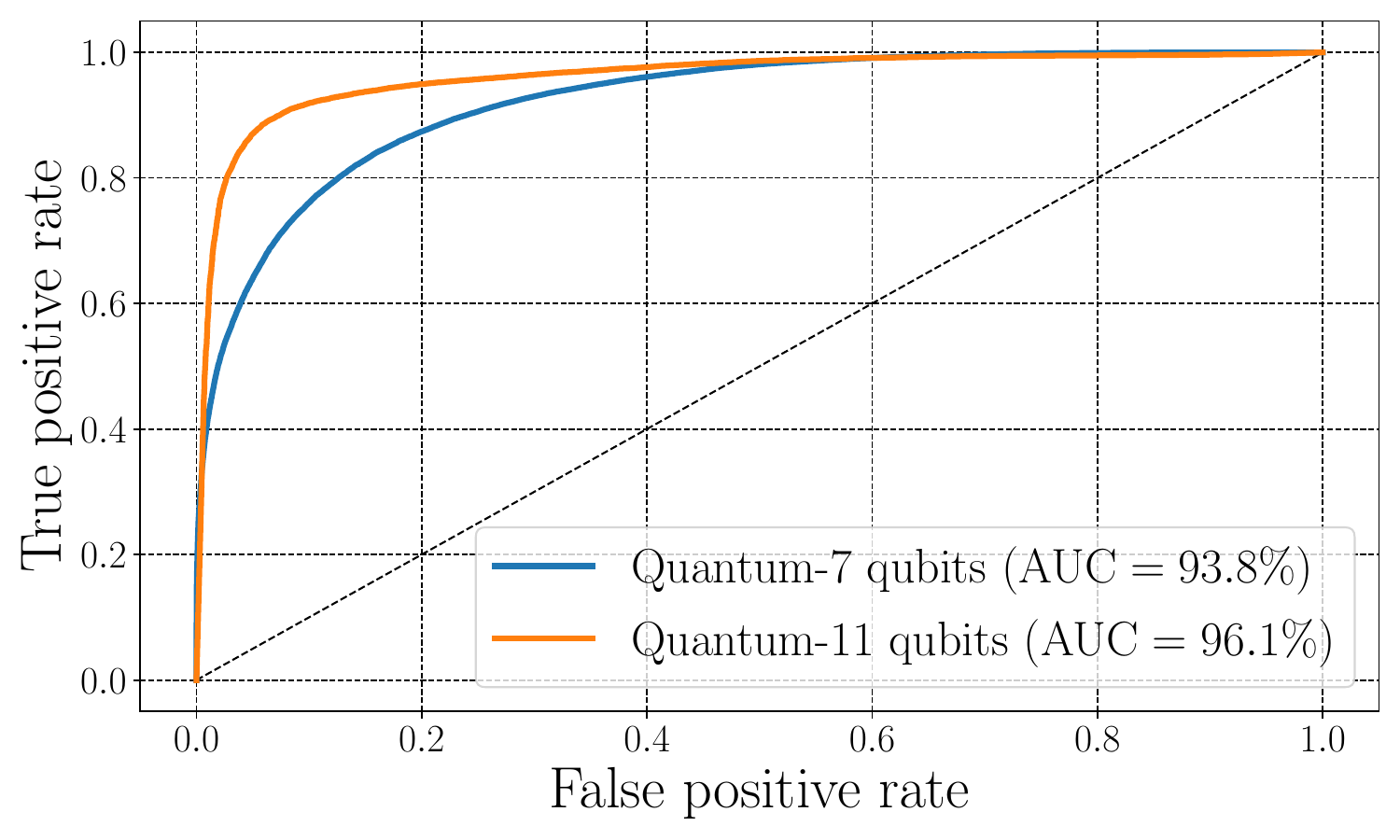}
    \caption{ROC curve for the used hybrid classical-quantum networks, with 7 and 11 qubits, tested on $10,000$ events. These networks are trained and evaluated using a classical emulator without any noise consideration.  }
    \label{fig:4}
\end{figure}

\begin{figure*}[ht!]
    \includegraphics[width=0.9\textwidth]{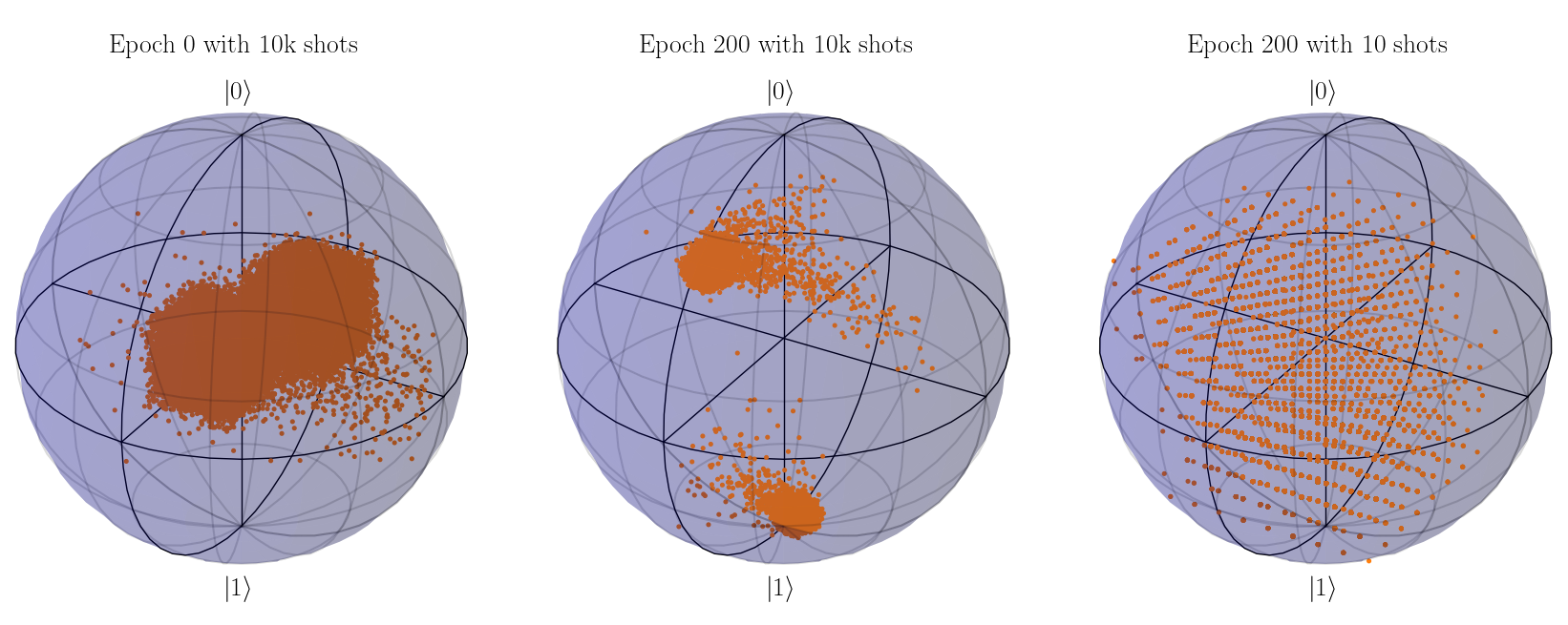} 
    \caption{Bloch sphere representation of the ancillary qubit with $20,000$ projected training events (orange points). Left: distribution of training events at the initial training step. Middle: distribution after 200 epochs of training with $10,000$ measurement shots, where signal events are distinctly separated from the background. Right: distribution after 200 epochs of training with only 10 measurement shots, where noise from the limited number of shots increases the overlap between signal and background events. In the right plot, the distinct positions of the two clusters diminished due to noise from the few-shot measurements. The Bloch sphere simulation is performed using the QuTiP package \cite{Johansson:2011jer}. }
    \label{fig:5}
\end{figure*}

Figure \ref{fig:4} shows the ROC curve for the hybrid classical-quantum networks for a test sample with $10,000$  events. The  hybrid network with $11$ qubits (green distribution) has the superior performance 
over the network with $7$ qubits (orange distribution).
This is expected as the expressive power of the network increase with the number of the qubits. 
The Area Under the Curve (AUC) quantifies the accuracy of the network prediction, and it is used to evaluate the performance of the networks. 
Specifically, the AUC for the quantum networks, with 7 and 11 qubits, are $93.8\%$ and $96.1\%$, respectively. By comparison, the AUC for the classical SL network is $95.0\%$. 

The measurement results for quantum circuits are subject to shot noise, 
which refers to the statistical fluctuations with the limited number of measurements. 
In each shot, the quantum state is projected onto one of the computational bases, based on the probability distribution determined by the wave function. Increasing the number of shots improves the accuracy of the measurement. 

The impact of shot noise is illustrated in Figure \ref{fig:5}. The left plot shows the Bloch-sphere representation of the ancillary qubit with initial training events in which the weights are randomly initialized.
The middle plot shows the distribution of training events on the ancillary qubit after $200$ epochs of training with $10,000$ shots, where the two classes are clearly separated. In contrast, the right plot demonstrates the effect of network training for the same number of epochs but with only $10$ shots. Here, shot noise blurs the distribution of each class, resulting in a significant overlap between the two classes.
This severely degrades the classification performance of the network when using a fidelity-based classifier. Specifically, 
the AUC is $79.0\%$ with $10$ shots,  compared with $96.1\%$ with $10,000$ shots.

\subsection{Refining few-shot measurement performance}%

To mitigate the reduction in network performance caused by shot noise for the small number of measurements, 
we replace the fidelity classifier with a clustering algorithm operating in the Hilbert space of the ancillary qubit. This can be achieved by utilizing a single classical Transformer encoder alongside the VQC, substituting the CSWAP test gates with simultaneous measurements along the $X, Y$ and $Z$ axes of the Bloch sphere. These measurements allow us to identify the position of each projected event on the Bloch sphere. By employing a clustering algorithm, we can categorize the projected events into two classes, signal and background. The performance of the network is evaluated based on the proximity of each test event to the centre of the two clusters.     

Hierarchical Density-Based Spatial Clustering of Applications with Noise (HDBSCAN) is applied to cluster the measured states of the training events. It has been introduced in \cite{10.1007/978-3-642-37456-2_14}  and was first used for jet clustering in \cite{Hammad:2024cae}. 

At its core, HDBSCAN works by first calculating a core distance for each data point on the Bloch sphere, which reflects the local density around that point by measuring its distance to nearby points. Using these core distances, it computes a mutual reachability distance between each pairs of points. This modified distance metric takes the local density into account, ensuring that points in dense regions are treated differently than those in sparser areas. 
Next, HDBSCAN constructs a minimum spanning tree from these mutual reachability distances, connecting all points in the dataset. This tree helps organize the data into a hierarchical structure, where clusters emerge as regions of high density are identified. By progressively removing the largest edges in the minimum spanning tree, the algorithm forms clusters at different density levels, creating a hierarchical tree of clusters. The final step involves analyzing this tree to find the most stable clusters. The algorithm selects these stable clusters as the final output. Points that do not belong to any cluster are classified as noise, and this is an important feature for the anomaly detection task.

\begin{figure}[th!]
    \includegraphics[width=0.48\textwidth]{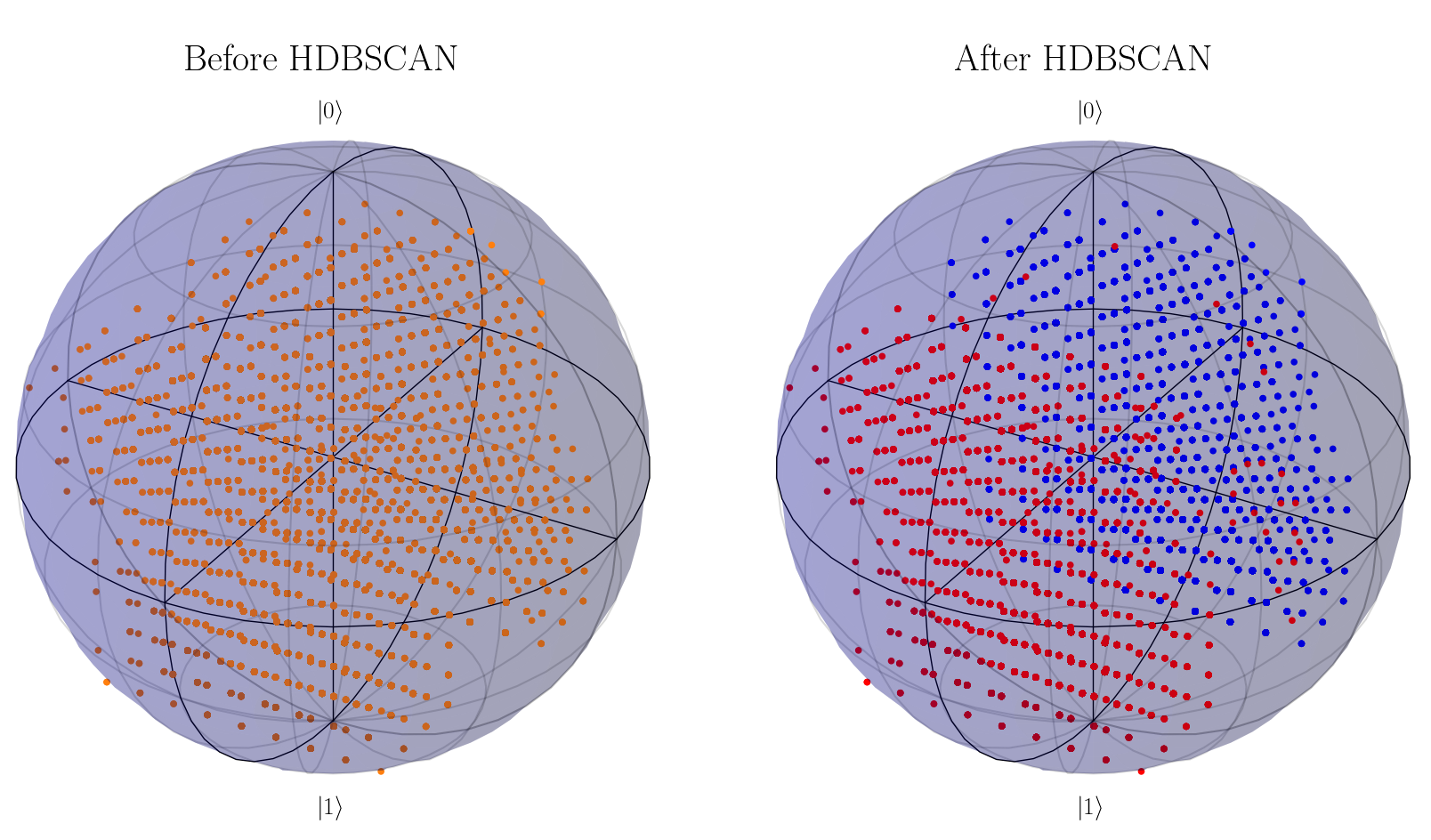}
    \caption{Bloch sphere representation of the training events is shown with 10 shots before applying HDBSCAN for clustering (left) and after clustering with HDBSCAN (right). Blue and red points represent the projected events from each cluster, as determined by the HDBSCAN results. Simulation of the Bloch sphere representation is done using QuTiP package \cite{Johansson:2011jer}.}
    \label{fig:6}
\end{figure}

The results of the HDBSCAN clustering are illustrated in Figure \ref{fig:6}. The left plot represents the projected training events on the ancillary qubit when the measurement is performed using  $10$ shots. HDBSCAN is used to cluster these events in two clusters\footnote{HDBSCAN automatically determines the number of clusters. We adjusted the algorithm's internal parameters to result in two clusters. Specifically, the minimum number of points per cluster was set to half the size of the training dataset.}. The right plot shows the result of HDBSCAN clustering, with the training events grouped into two classes: signal and background. The centre of each cluster was computed, and the network performance was evaluated based on the proximity of test events to the cluster centres. As a result, the network performance improved from an AUC of $79.0\%$, using the fidelity classifier, to $89.7\%$ after applying HDBSCAN. While this discussion focuses on mitigating shot noise, this method can also be applied to address other sources of noise, making it suitable for tests on real quantum computers. 

\section{Conclusion}
\label{sec:5}

In this paper, we apply SL for anomaly detection searches for a heavy scalar resonance decaying to the di-Higgs boson at the LHC. We adopt two similarity learning networks, classical and hybrid classical-quantum. Both networks comprise a pair of Transformer encoders that share their weights. The role of the transformer encoder is to map the high-dimensional input data into a lower-dimension latent space. 
A projection head is added to the latent space data to compute the contrastive loss function, while for the hybrid classical-quantum network, the latent data is mapped onto VQC exploiting the power of quantum computers to express high complex data structure. 
The similarity of the input pair is then computed in terms of the overlap between the measured quantum state using CSWAP test gates. A metric loss function, the HS loss function, is used to minimize the overlap between the measured quantum state of the events from positive pairs and maximize the overlap between events from negative pairs. 

To demonstrate the capability of the VQC with a larger number of qubits, we explore two hybrid classical-quantum networks with $7$ and $11$ qubits. In the ideal case, where quantum noise is ignored, the hybrid network with $11$ qubits outperforms the other networks. However, when quantum noise from a limited number of measurement shots is introduced, the performance degrades significantly. To address this, we utilize the qubit Hilbert space to recluster the mapped events from different classes. By applying HDBSCAN, a density based clustering algorithm,  the network performance improved by $9\%$ over the fidelity classifier test. While this technique effectively mitigates shot noise, it can also enhance performance in the presence of other quantum noise sources that affect the measurement of the quantum fidelity.

Although SL can efficiently be adjusted for anomaly-detection tasks, a challenge lies in selecting negative pairs. In practice, negative samples are often assumed to be dissimilar by default, but this may not always hold true. Finding more informative or hard negative samples, those that are difficult to distinguish from positive ones can greatly improve the performance of the model; identifying these samples, however, remains a challenge. Another challenge is that SL  requires large batch sizes to store negative samples, which is computationally expensive.

Finally, while current quantum devices are limited, fault-tolerant quantum computing promises to overcome these limitations in the future. Simulating quantum machine learning methods prepares for the eventual deployment of quantum algorithms on fault-tolerant quantum computers. By simulating the behaviour of quantum algorithms in ideal conditions, we can predict their performance and scalability, laying the groundwork for future breakthroughs. The LHC data analysis offers the best environment for testing the validity of quantum algorithms with the current NISQ computers and the scalability for fault-tolerant quantum devices.

\section*{Acknowledgments}

This work is partially funded by grants 22H05113, “Foundation of Machine Learning Physics,” Grant in Aid for Transformative Research Areas, and 22K03626, Grant-in-Aid for Scientific Research (C).
This work is also supported in part by JSPS Grant-in-Aid for Scientific
Research (Nos. 20H05860, 23H01168), and JST, Japan (PRESTO Grant No.~JPMJPR225A, Moonshot R\&~D Grant No.~JPMJMS2061).
This work was performed in part at the Aspen Center for Physics, which is supported by National Science Foundation grant PHY-2210452.

\appendix
\section{Basics of VQC training}
\label{App:A}
Quantum computers can carry out supervised and unsupervised learning by leveraging parameterized quantum circuits, which act as models that map input data to predictions. These circuits, known as VQCs, extract features from classical data by first encoding them into quantum states. After the data is encoded, unitary transformations are applied to these quantum states, and the results are measured by projecting them onto the computational basis. The encoding method of classical data plays a crucial role in determining the expressive power of the VQC.

In applying variational methods to approximate the eigenvalues and eigenvectors of the Hamiltonian on a quantum computer, the process can be simplified into three main stages that define the algorithm's workflow:

\begin{itemize}
    \item Prepare the states by encoding the classical data onto quantum computers, with some trainable parameters.
    \item Measure the expectation values from the readout qubit.
    \item Use some classical optimizer to obtain new values for the trainable such that the measured expectation values are closer to the prediction value. 
    \item  Iterate this procedure until a loss function is minimized.
\end{itemize}

The parameterization of the data encoding can influence the decision boundaries of the final predictions, and should therefore be selected to best suit the problem at hand \cite{Farhi:2018nhu,Blance:2020nhl}. There are several types of parameterizations, including basis encoding, amplitude encoding, and angle encoding. These methods map input data to qubits based on fixed rotations, while another approach is to train the data encoding to maximize the Hilbert space distance between different inputs \cite{perez2020data,nghiem2020unified}. A more recent technique involves repeated and incremental encoding \cite{Periyasamy:2022lzp}, which maps high-dimensional classical data into a Variational Quantum Circuit (VQC) using fewer qubits. For angle encoding, a common choice is to use single-qubit Pauli rotation gates, $R_\alpha(x) = e^{-ix\sigma_\alpha/2}$, where $\sigma_\alpha$ ($\alpha=X, Y,Z$) are the three Pauli matrices.
Pauli rotation gates are single qubit gates that encode a single classical input to a single qubit. One can instead use a more generic rotation to encode three features into a single qubit via sequential rotations as \cite{Barenco:1995na}
\begin{equation}
  \begin{split}
          U &= e^{i\delta}R_Z(\alpha)R_Y(\beta)R_Z(\gamma)\\ \hspace{6mm} &= \left ( \begin{array}{cc}
       e^{i(\delta+(\alpha+\gamma)/2)} c_{\beta}  & -e^{i(\delta-(\alpha-\gamma)/2)} s_{\beta}  \\
       e^{i(\delta+(\alpha-\gamma)/2)} s_{\beta} & e^{i(\delta-(\alpha+\gamma)/2)} c_{\beta}
    \end{array} \right )  \,, \end{split}
\end{equation}
where $c_{\beta}=\cos({\beta}/{2})$ and $s_{\beta}=\sin({\beta}/{2})$. The global phase $e^{i\delta}=\pm 1$ is chosen such that the determinant of $U$ is 1.  From these parameter definitions, we can utilize a maximum of three input dimensions per unitary operation \cite{ozdenizci2019adversarial,9415627}. After the quantum states are prepared, the VQC  maps the prepared state to another state via a set of unitary transformations, $|\psi(x,\theta)\rangle = U(\Vec{\theta})|x\rangle$, with $\Vec{\theta}$ are the tunable parameters to minimize the error between the model predictions and the true values. In general, the unitary transformations of the prepared states can be decomposed into a series of sequential unitary gates as
\begin{equation}
    U(\Vec{\theta}) = U_j(\Vec{\theta}_\sigma) \dots U_3(\Vec{\theta}_\gamma) U_2(\Vec{\theta}_\beta) U_1(\Vec{\theta}_\alpha) \,,
\end{equation}
with $j$ representing the maximum number of unitary gates in the quantum layer. For circuits with a larger number of qubits, $U$ is composed of unitary rotation gates and entangling gates. Common two-qubit entangling gates include Controlled-NOT (CNOT) and Controlled-Z gates, which do not have tunable parameters. These gates flip the state of one qubit based on the value of a control qubit. Increasing the number of entangled qubits in the circuit takes advantage of interference effects, enabling the input data to be mapped to a higher dimensional space and allowing learning from small datasets \cite{Sim:2019yyv,Guan:2020bdl,Caro:2021mgf}. After all unitary operations are applied, the expectation value of the quantum state can be measured on one of the computational basis states using the Pauli-Z operator on the readout qubit as
\begin{equation}
    f_\theta(x) = \langle\psi(x,\theta)|\sigma_Z| \psi(x,\theta)\rangle \,,   
\end{equation}
where $\sigma_Z$ is the Pauli-Z operator.
Optimization of the quantum circuit parameters can be done by minimizing a loss function between the eigenvalues of the measured quantum state and the true labels. Here, as in the classical machine learning methods, minimization of the loss function can be done using classical optimizers such as the Adam optimizer. The backpropagation is computed using the parameter shift rule \cite{PhysRevA.98.032309,PhysRevA.99.032331} as 
\begin{equation}
    \frac{\partial}{\partial\theta} \mathcal{F} = \frac{\mathcal{F}(\theta+\delta)-\mathcal{F}(\theta-\delta)}{2} \,,
\end{equation}
where $\mathcal{F}$ is the quantum circuit and $\delta= {\pi}/{2}$. After optimizing the parameters of the quantum circuit, network performance can be evaluated on a new unseen dataset.

\section{The role of the  Transformer encoders}%
\label{subsec:1.3}
In an SL task, the encoders pair plays a pivotal role by facilitating the simultaneous processing of the  input pairs. In this framework, two identical neural-network encoders with shared weights are employed to project the input pairs onto low dimensional space.
It is crucial to share the weights across the two encoders when comparing the embedding from different augmentations of the same input. By enforcing the same weights, the two encoders have a shared understanding of the features that constitute similarity. This specific structure of the two encoders is independent of the type of the neural network or the structure of the input data. Accordingly, different types of neural network encoders can be used, e.g.\ multi-layer perceptron, graph neural network or attention-based Transformer encoder. 

Transformers have recently gained attention for particle cloud analysis at the LHC due to thier ability to model complex and high dimensional data. The motivation to apply transformer encoders to particle clouds stems from their inherent ability to model interactions between particles irrespective of their spatial proximity \cite{Komiske:2017aww,Qu:2019gqs}. This can be guaranteed by structuring the input data in which events are represented as unordered sets of particles where each particle is characterized by its low level information, e.g.\ four momenta and spatial coordinates. The core of  Transformer layers is the attention mechanism that enables the model to focus selectively on different parts of the input sequence. In general, the attention mechanism operates by assigning different weights to different elements in the input sequence, emphasizing the more relevant parts while downplaying the less relevant ones.  

Considering the input dataset as $X_{i,j}$ with $i,j$ representing the particle and feature tokens, respectively. A fully connected linear layer is used to generate tunable matrices as 
\begin{gather*} 
    Q_{i,d} = X_{i,j}\cdot W^Q_{j,d}, \hspace{6mm} K_{i,d} = X_{i,j}\cdot W^K_{j,d},\\
    V_{i,j} = X_{i,j}\cdot W^V_{j,j}\,,
    \end{gather*} 
where $Q,K$ and $V$ are the query, key and value matrices, respectively, $W$ are the tunable matrices added by the fully-connected linear layer, and the dimension $d$ maps the inputs to higher dimensions. The scaled dot attention score can be computed as 
\begin{equation} 
    \alpha_{i,i} = \textrm{softmax}\left(\frac{Q_{i,d} \cdot K^T_{i,d}}{\sqrt{N}} \right)\,,
\end{equation}
where $N$ is the size of the input data and the normalization factor $\sqrt{N}$ is added to avoid exploding gradients.  Attention output is computed by multiplying the attention score by the $V$ matrix as 
\begin{equation}\label{ZV}
    \mathcal{Z}_{i,j} = \alpha_{i,i}\cdot V_{i,j}\,.
\end{equation}
The attention of Eq.~\eqref{ZV} describes a single attention head.  Multiple parallel attention heads are employed for our analysis:
\begin{equation}
    \mathcal{O}_{i,j} = \textrm{concatenate} \left(\mathcal{Z}^1_{i,j},\mathcal{Z}^2_{i,j} \ldots \mathcal{Z}^n_{i,j} \right)\cdot W_{n\ast j, j}\,,
\end{equation}
with $W_{n\ast j, j}$ being the learnable linear transformation matrix to retain the dimensions of the input. The computed attention is then used to scale the input  via a skip connection as
\begin{equation}
\widetilde{\mathcal{X}}_{i,j} = \mathcal{O}_{i,j} + X_{i,j}\,.
\end{equation}
The transformed dataset $\widetilde{\mathcal{X}}_{i,j}$ highlights the importance of individual particle tokens in the event in the network classification decision. Additionally, the transformed dataset has the same dimensions as the input dataset, therefore the multi-head attention layer can be applied repeatedly.  
A final normalization fully connected projection layer is added to map the attention output into a single vector that can be passed to the contrastive loss or to be embedded onto the VQC.

\bibliographystyle{JHEP}
\bibliography{biblo}
\end{document}